\documentclass[10pt,aps,superscriptaddress]{revtex4}
\usepackage{amsmath}
\usepackage{mathpazo}
\newcommand{\bc}{\begin{center}}
\newcommand{\ec}{\end{center}}
\newcommand{\tr}{\mathop{\rm tr}\nolimits}
\begin{document}
\title{Classical solutions for Yang-Mills-Chern-Simons field coupled to an external source}
\author{Vivek M. Vyas}
\email{vivek@prl.res.in}
\affiliation{ IISER Kolkata, Mohanpur Campus, PO: BCKV Campus Main Office, Mohanpur, West Bengal, India-741 252.}
\author{T. Soloman Raju}
\affiliation{Birla Institute of Technology \& Science-Pilani, Goa Campus, Zuarinagar, Goa, India-403 726.}
\author{T. Shreecharan}
\email{shreet@imsc.res.in}
\affiliation{The Institute of Mathematical Sciences, C.I.T.Campus,Taramani, Chennai, Tamil Nadu, India-600 113.}
\begin{abstract}
\bc
\small{Abstract}
\ec
We find wide class of exact solutions of Yang-Mills-Chern-Simons theory coupled to an external source, in terms of doubly periodic Jacobi elliptic functions. The  obtained solutions include localized solitons, trigonometric solutions, pure cnoidal waves, and singular solutions in certain parameter range. Furthermore, it is observed that these solutions exist over a nonzero background.
\end{abstract}
\maketitle
\section{Introduction}

Abelian and non-Abelian gauge theories in lower space-time dimensions manifest in diverse condensed matter systems. The planar $2+1$ dimensional field theories find extensive applications in the description of quantum Hall effect \cite{wen}, high-$T_{c}$ superconductivity \cite{bhask} and recently in graphene \cite{castro}. Both Abelian and non-Abelian gauge symmetries often emerge in constrained systems e.g., deviations from half filling in the t-J model \cite{zouander,pkpasim}. The gauge fields, initially appearing as Lagrange multipliers, acquire dynamics through their coupling to the fermionic degrees of freedom via quantum corrections. Unlike even space-time dimensions, in $(2+1)$ dimensional world, the dominant term in the low energy effective action happens to be the topological Chern-Simons (CS) term \cite{jackiw,redlich,pkpdas}. Being parity and time reversal symmetry violating, this term naturally appears when the underlying fermionic systems break these symmetries \cite{redlich,pkpdas}. In presence of this term, the fermions can acquire additional statistical phase, thereby becoming anyons \cite{wilzee,wilczek}. These theories have been systematically investigated since last thirty years, yielding significant connections with conformal field theories and knot theory \cite{polyakov,witten1, witten2}. 

Non-Abelian CS field coupled to a matter field has been well studied, and it has been shown that this system is connected with a host of integrable models \cite{dunnejackiw}. The next important contribution to the effective action is the two derivative kinetic term, which gives dynamics to the underlying gauge field. Simplest example of such a theory is the Maxwell-CS theory, in which a usual Maxwell Lagrangian is augumented with a CS term. It is known that in such a theory, photon acquires a gauge invariant mass \cite{dunne}. In the presence of interacting matter, this theory is known to have charged vortices \cite{khare1} and anyonic excitation \cite{dunne}. Interestingly, this system can also show chaotic dynamics under suitable conditions \cite{bambah1, bambah2}. The fact that the CS theory coupled to matter system has rich dynamical structure, from integrability to chaotic behaviour, makes this systems interesting for analysis. 

In the present paper, we investigate non-Abelian SU(2) CS theory coupled to external matter field and investigate the structure of the classical solution space. The initial interest in the study of classical solutions can be traced to the fact that for a non-Abelian gauge theory like QCD, the ground state is not known. It was hoped that a classical study might provide some insight. The study indeed has thrown up some interesting solutions like the Wu-Yang monopole \cite{Wu}, non-Abelian plane waves \cite{Coleman}, 't Hooft-Polyakov monopole \cite{Hooft}, instanton \cite{Polyakov} and meron \cite{Furlan} solutions. The equations of motion for non-Abelian gauge theories in general are complicated non-linear partial differential equations (PDEs), and hence have a rich solution space.

The fact that equations of motion are coupled nonlinear PDEs, which unlike linear ones, do not always exhibit exact solubility. By employing an ansatz we are able to map this infinite dimensional dynamical system to finite dimensional ordinary DE. Making use of a conformal transformation, we show that this finite dimensional problem is exactly soluble and find a wide variety of solutions from periodic to localised soliton solutions. The solution show a signficantly different asymptotic behaviour than the solutions that have been studied so far.

The manuscript is organized as follows. In the subsequent section, we present the action and the equations of motion of the gauge field under consideration. In Section III, we analyse the low energy dynamics offered by these classical equations. We summarize our results with a discussion in Section IV. 

\section{The equations of motion}

The Lagrangian for the SU(2) Yang-Mills-Chern-Simons theory (YMCS) is given by
\begin{eqnarray} \label{ymcslag}
{\cal L}_{YMCS} = -
\frac{1}{4} F_{\mu \nu}^{a} F^{\mu \nu {a}} +
\frac{M}{4}\epsilon^{\mu\nu\rho} [F_{\mu\nu}^a A_{\rho}^a -
\frac{g}{3}\epsilon^{abc} A_{\mu}^{a}A_{\nu}^{b}A_{\rho}^{c}] ,
\end{eqnarray}
where $F_{\mu\nu}^a =
\partial_{\mu}A_{\nu}^a - \partial_{\nu}A_{\mu}^a +
g\epsilon^{ abc}A_{\mu}^bA_{\nu}^c $
is the field strength tensor, $M$ is the topological mass of the gauge field $A$, and $g$ is the coupling constant. The $\epsilon$'s are the completely antisymmetric structure constants of the gauge group. 

The above Lagrangian (\ref{ymcslag}), is not explicitly gauge invariant, since it changes by a full
derivative under a gauge transformation
\begin{eqnarray}
S_{YMCS} = \int d^3 x
\, \mathcal{L}_{YMCS} \longrightarrow \int d^3 x \, \mathcal{L}_{YMCS} + \frac{8 \pi^2 M}{g^2}
W (U) ,
\end{eqnarray}
where $ W (U)$ is the winding number:
\begin{eqnarray}
W (U) = \frac{1}{24\pi^2} \int d^3x \epsilon^{\alpha\beta\gamma} \tr
[\partial_\alpha UU^{-1}\partial_{\beta} UU^{-1}\partial_{\gamma}
UU^{-1}].
\nonumber
\end{eqnarray}
Here $U(x)=\exp [i\sigma^a \lambda^a(x) ]$ is the non-Abelian gauge transformation and $\sigma^a$ are the the Pauli matrices. Demanding the gauge invariance of the partition function, results in the quantization of the CS parameter $M$:
\begin{eqnarray} 
M = \frac{g^2 n}{4\pi} , \quad n \in Z.
\end{eqnarray}

The YMCS gauge field can be coupled to an external source field by augumenting Lagrangian  ${\mathcal{L}_{YMCS}}$ with an interaction term: 
\begin{equation}
\label{source}
\mathcal{L}_S = A_\mu^a J^{\mu \, a}, 
\end{equation}
to give the classical Euler-Lagrange equation:
\begin{eqnarray}\label{eof}
D_{\nu}F^{\nu\mu a} +
\frac {M}{2} \epsilon^{\mu \nu \rho} F_{\nu \rho}^a = J^{\mu \, a} ;
\end{eqnarray}
where $ D_{\mu} F_{\nu \rho}^{a}=\partial _{\mu} F_{\nu \rho}^a + g\epsilon^{abc}
A_{\mu}^b F_{\nu \rho}^c $. 

Equations (\ref{eof}) are coupled nonlinear PDEs and have infinite degrees of freedom. Ideally, it is desirable to have exact solutions to these equations, so that a complete  understanding of the classical phase space is obtained. However, there are only a handful such theories, which are said to be {\emph{integrable}}, for which the classical phase is completely understood, Sine-Gordon model being one of the celebrated ones \cite{dasintegrable}. In such an integrable theory, there exists an infinite dimensional symmetry algebra, which provides infinite conserved quantities, which leads to exact solubility of the equations of motion.  

The symmetry algebra of Yang-Mills-Chern-Simons theory is only finite dimensional, and hence does not belong to the integrable class, like most non-trivial realistic theories. Hence, the hope to solve Eq. (\ref{eof}) in its full generality, is futile. However, the nonintegrable set of equations open up an interesting possibility of having chaos, wherein the evolution of the system shows sensitive dependence on initial conditions, and evolution from any two infinitesimally close points in the phase leads to exponential separation in the trajectories \cite{laksh}. It is obvious that, such a behaviour can not be described analytically, and numerical evolution methods are used to study such systems. Still, the equations (\ref{eof}) are too difficult to study right away, even numerically. A convenient way to sneak in the dynamics, is to reduce the degrees of freedom, such that the system then becomes tractable. The dynamical systems with two and three dimensional phase space are well understood \cite{laksh}. By choosing a convenient ansatz, one can reduce the dimensionality of phase space, and then study the dynamics of the resultant system. Although such a study would be restrictive, but it would certainly throw some light on the dynamics of the parent model.

\section{Time dependent solution}

The classical solutions of YMCS field theories with and without a source term have been investigated over the years by a number of authors \cite{bambah1,bambah2,oh1,Wudka,oh2,oh3,rogal}. Restricting ourselves to purely time dependent situations of (\ref{eof}): $ A_{\mu}^a = A_{\mu}^a (t)$ and $J^{\mu \, a} = J^{\mu \, a}(t)$, we choose the following ansatz \cite{rogal} for the fields, along with external source as:
\begin{equation}
A_{\mu}^a = \delta_{\mu}^a f_{a}(t), \quad A^{\mu a} = g^{\mu a}f_{a}(t), \quad J^{\mu a} = g^{\mu a}j_{a}(t),
\end{equation}
where $ a=1,2,3, \, \mu =1,2,3\ (x_1=t , x_2 = x , x_3 = y)$ and $g_{\mu \nu} = {\mbox {diag}}(+,-,-)$. Note that there is no summation for $a$. Although we are restricting to purely time dependent gauge field configurations, our theory is Lorentz covariant. And this symmetry permits us always to boost our solutions to a frame which is moving with some non-zero relative velocity. Hence, the solutions that we will be discussing henceforth can well be thought of as wave packets with a functional form: $f(k_{\mu} x^{\mu})$. With abovementioned ansatz, the equations of motion can be written in explicit component form as
\begin{eqnarray}
M g f_2 f_3 + g^2 f_1(f_2^2 +f_3^2) = j_1 , \nonumber \\
M g f_1 f_3 -\ddot{f_2} + g^2 f_2(f_1^2 -f_3^2) = j_2 , \nonumber \\
M g f_1 f_2 -\ddot{f_3} + g^2 f_3(f_1^2 -f_2^2) = j_3 ,\\
M \dot{f_2} + g(\dot{f_1}f_3 + 2f_1\dot{f_3}) = -j_2 , \nonumber \\
M \dot{f_3} + g( \dot{f_1}f_2 + 2f_1\dot{f_2}) = -j_3 , \nonumber
\end{eqnarray}
where $\dot{f_{a}}$ denotes derivative with respect to time. Notice, that this system has only finite number of degrees of freedom, and hence finite dimensional phase space.

Our goal of solving nonlinear PDEs (\ref{eof}) is reduced to finding non-trivial solutions for the above system of coupled ordinary differential equations. To this end let us set $f_1= \kappa$ (a constant) and take $f_2=f_3=f(t)$. With $j_1=0$ the first equation in the above set gives $\kappa = -M/2g$. Now we set $j_2=j=j_3=j$ to get
\begin{equation} \label{elliptic}
\ddot{f}(t) + \frac{M^{2}}{4} f(t) + g^2 {f(t)}^{3} + j(t) = 0,
\end{equation}
which when solved would give information about dynamics of this model. When $j(t)$ is absent this equation reduces to an elliptic differential equation, whose solutions can be expressed in terms of elliptic functions.
Unlike linear differential equations, the above equation does not allow general solution for any arbitrary function $j$. Instead, only special solutions for certain form of $j$ can be found. At this point, we emphasize that for propagating solitons (in co-moving frame), the equations of motion of YMCS field coupled to an external source, yield nonlinear Scrodinger equation interacting with an external source. Inspired by this fact, we investigate the localized solutions of Eq. (\ref{elliptic}) for $j(t)=j$ \cite{soloman1,soloman2}.

Further, in this case, solution to above equation is related to Jacobi elliptic functions via a conformal (M\"obius) transformation, which amounts to say that:
\begin{equation} \label{ft}
f(t) = \frac{A + B \ \phi(t,m)}{C + D \ \phi(t,m)}.
\end{equation}
Here $A, B, C$ \& $D$ are real constants, such that $AD-BC\neq0$, and $\phi(t,m)$ is one of the twelve Jacobi elliptic functions, which satisfy a differential equation of the kind:
\begin{equation} \label{firstint}
\ddot{\phi} + a \phi + b \phi^{3} = 0,
\end{equation}
with $a$ \& $b$ being real constants. These Jacobi elliptic functions are generalization of circular and hyperbolic trigonometric functions \cite{fokas}. Each of these functions, is characterised by a modulus parameter $m$, for example, the elliptic function $cn(x,m)$ goes to $cos(x)$ in the limit $m\rightarrow0$, and to $sech(x)$ when $m\rightarrow1$. Although for $m>1$ these elliptic functions are defined, they can be rewritten in terms of elliptic functions with $0<m<1$. The first integral of Eq. (\ref{firstint}) is
\begin{equation} 
E_{0} = \frac{{\dot{\phi}}^{2}}{2} - \frac{a\phi^{2}}{2} - \frac{b\phi^{4}}{4},
\end{equation} 
and substituting (\ref{ft}) in equation (\ref{elliptic}), one ends up to get consistency conditions in equation parameters and the solution parameters:
\begin{align} \label{eq12}
& -4 B C D E_{0}+ 4 A D^2 E_{0} + \frac{M^2}{4} A C^2 + g^2 A^3 + j C^3 = 0,\\
& a B C^2 - a A C D + \frac{M^2}{4} B C^2 +  \frac{M^2}{2} A C D + 3 g^2 A^2 B + 3 j C^2 D = 0,\\
& -a B C D + a A D^2 + \frac{M^2}{2} B C D + \frac{M^2}{4} A D^2 + 3 g^2 A B^2 + 3 j C D^2 = 0,\\ \label{eq15}
& b B C^2 - b A C D + \frac{M^2}{4} B D^2 + g^2 B^3 + j D^3 = 0.
\end{align}
Equations (\ref{eq12}) to (\ref{eq15}) can be solved in terms of solution parameters and equation parameters to give:
\begin{align}
& E_{0}=-\frac{b A C^3+a B C^2 D+a A C D^2}{4 B D^3},\\
& M^{2}=4\frac{-3 b A C^2 - 2 a B C D - a A D^2}{D (-B C+A D)},\\
& g^2=\frac{-b C^3-a C D^2}{B (B C-A D)},\\
& j=-\frac{b A B C^2 + b A^2 C D + a B^2 C D + a A B D^2}{D^2 (B C-A D)}.
\end{align}
By choosing various values for $A, B, C$ \& $D$, we get a spectrum of values for the equation parameters for which above considered ansatz is a solution. For example, for $\phi=cn(t,m)$, $a=2m-1$ and $b=-2m$, and above relations with these values determine the equation parameters for which this ansatz is a solution. 
Interestingly, for $m=1$, this choice of ansatz yields dark and bright localized solitons, which are of the type:
\begin{equation}
f = \frac{A + B \ Sech(t)}{C + D \ Sech(t)}.
\end{equation}
When $AD>BC$ one gets dark solitons, and when $AD<BC$ one gets bright solitons (and ofcourse if $AD=BC$ then there is only trivial solution), and for $C=1$ and $D=-2$, $f(t)$ will become a singular solution. On the other hand, when $m=0$, one gets periodic solutions of the kind:
\begin{equation}
f = \frac{A + B \ Cos(t)}{C + D \ Cos(t)}.
\end{equation}
Another interesting class of explicit solutions of this dynamical system emerges by choosing $\phi=sn(t,m)$. This case, in the limit of elliptic modulus m->1 yields kink-type of solution. Hence, various possible choice of elliptic functions and possible values for $A$, $B$, $C$ \& $D$, which ensure the positivity of $M^{2}$ and $g^2$, yield various valid solutions to equation (\ref{elliptic}). An interesting point to notice is that, the solutions are necessarily defined over a constant background $\frac{A}{C}$, which is present as a consequence of a constant source term. Hence, the solutions found above do not obey equation (\ref{eof}) with the usually assumed boundary conditions $A^{\mu a} \rightarrow 0$ as $x,y\rightarrow \pm \infty$.

One may wonder, whether this model (\ref{elliptic}) exhibits chaos or not, for some nontrivial $j(t)$. For, $j(t)=Cos(t)$, equation (\ref{elliptic}) resembles the famous forced Duffing oscillator, which is a standard example to demonstrate chaotic dynamics. Unfortunately, positivity of $M^2$ and $g^2$, does not allow this model to exhibit chaos, which requires any one of these to be negative. Hence, we conclude that this model, defined by (\ref{elliptic}), can not be made chaotic by coupling it with an external source.

\section{Conclusion}

We have shown that Yang-Mills Chern-Simons theory coupled to a constant external current density, yields a host of interesting solutions, from localised soliton type to periodic solutions of infinite extent. The solutions are necessarily defined over a nonzero constant background. It has been shown that this model, to the extent of validity of the ansatz, can not exhibit chaos. It worth mentioning here that if the sources are not external, but say are dynamic like a Higgs field, then such a system indeed exhibits chaos \cite{bambah1,bambah2,oh3,rogal}. It is interesting to speculate about the quantum theory defined over above dicussed ground states. We think such a study would certainly be relevant for understanding nonperturbative aspects of low dimensional gauge theories, specially it may be relevant for quantum Hall effect physics. Finally, we think that the technique employed here may be helpful in understanding the semiclassical and nonperturbative dynamics of other gauge theories. \\

\textit{Acknowledgements:-} We thank Prof. Prasanta K. Panigrahi for his careful reading of this manscript and many insightful discussions. The major part of this work was done while VMV was in Physical Research Laboratory, Ahmedabad; and he acknowledges moral and finanical support recieved from them.

\end{document}